\documentclass[aps,showpacs,11pt]{revtex4}
\usepackage{dcolumn}
\usepackage{graphicx}
\usepackage{amsmath}
\usepackage{amsfonts}
\usepackage{amssymb}
\usepackage{psfrag}
\usepackage{wrapfig}
\usepackage{subfigure}
\usepackage{makeidx}
\usepackage{bm}
\usepackage{epsf}
\usepackage{hyperref}
\usepackage{color}
\usepackage{multirow,dcolumn}
\usepackage{graphicx}

\begin{document}

\title{Holographic entanglement entropy close to crossover/phase transition in strongly coupled systems}

\author{Shao-Jun Zhang}
\email{sjzhang84@hotmail.com}
\affiliation{Institute for Advanced Physics and Mathematics, Zhejiang University of Technology, Hangzhou 310023, China}
\date{\today}

\begin{abstract}
  We investigate the behavior of entanglement entropy in the holographic QCD model proposed by Gubser et al. By choosing suitable parameters of the scalar self-interaction potential, this model can exhibit various types of phase structures: crossover, first order and second order phase transitions. We use entanglement entropy to probe the crossover/phase transition, and find that it drops quickly/suddenly when the temperature approaches the critical point which can be seen as a signal of confinement. Moreover, the critical behavior of the entanglement entropy suggests that we may use it to characterize the corresponding phase structures.
\end{abstract}


\maketitle

\section{Introduction}

In the past two decades, AdS/CFT~\cite{Maldacena:1997re,Gubser:1998bc,Witten:1998qj} or the more generic gauge/gravity duality has attracted lots of attention and efforts, which relates a quantum field theory (QFT) in $(d+1)$ dimensions to some gravitational theory in $(d+2)$ dimensions. As a strong/weak duality, it provides a powerful tool to deal with strongly coupled field systems for which traditional methods of perturbative QFT confront great challenge or even break down. It has been applied on various areas of modern theoretical physics, including QCD~\cite{Mateos:2007ay,Gubser:2009md,CasalderreySolana:2011us}, condensed matter physics~\cite{Hartnoll:2009sz,Herzog:2009xv,McGreevy:2009xe,Horowitz:2010gk,Cai:2015cya} and cosmology~\cite{Banks:2004eb}, and achieved great successes.

On the other hand, experiments of heavy ions collisions on RHIC~\cite{Adams:2005dq,Back:2004je,Arsene:2004fa,Adcox:2004mh} have opened a novel window into the physics of strongly interacting hadronic matters. The existing data suggest the following evolution picture: After collision, the hot QCD matters undergo a very fast thermalization process to reach thermal equilibrium where a ball of quark-gluon plasma (QGP) forms; And subsequently, the QGP expands to cool down until the temperature falls below the QCD transition (or crossover) point where it finally hadronizes. The QGP can be described very well by relativistic hydrodynamics with a very small $\eta/s$~\cite{Luzum:2008cw}, where $\eta/s$ is the ratio of shear viscosity to entropy density. This implies that the QGP is strongly coupled, and thus a treatment beyond the perturbative QCD is called for. With the help of the most well-established example of the AdS/CFT correspondence, namely the duality between $N=4$ superconformal Yang-Mills (SYM) theory in four-dimensional Minkowski spacetime and type IIB supergravity in $AdS_5 \times S^5$, $\eta/s$ of $N=4$ SYM is produced which is very close to the hydrodynamic result~\cite{Policastro:2001yc,Kovtun:2004de}. This remarkable result shows the validity and powerfulness of the holographic method. However, $N=4$ SYM is a conformal field theory and thus does not exhibit crossover behavior or any kind of phase transition which is very different from QCD. Moreover, lattice data indicates that the QGP is not a fully conformal fluid in the relevant RHIC energy range $1 \leq T/T_c \leq 3$, and the deviation from conformality may play an important role near the crossover/phase transition~\cite{Karsch:2007jc,Meyer:2007dy,Meyer:2008sn}. Therefore, it is interesting to seek more realistic holographic models to model non-conformal field theories such as QCD.

Now there are various ways to construct holographic models dual to non-conformal field theories, either following a top-down approach by studying a specific gravitational theory which has a string theory construction or a bottom-up approach in which the gravitational background is phenomenologically fixed to fit the lattice QCD (lQCD) data. Recently, in Refs.~\cite{Gubser:2008ny,Gubser:2008yx} Gubser and his collaborators proposed an interesting bottom-up holographic model intending to mimic the equation of state of QCD. In their model, beyond the Einstein gravity sector, a nontrivial massive scalar field as well as a judicious choice of its self-interaction potential are introduced in the bulk to break the conformal symmetry. The scalar potential has several parameters. By choosing appropriate values of these parameters, the model can exhibit a crossover behavior at some critical temperature and the equation of state generated agrees well with the result from lQCD~\cite{Borsanyi:2012cr}. Moreover, this simple model can also realize various types of phase transitions by choosing other values of the parameters in the scalar potential, for example the first and second order phase transitions. Thus, in additional to mimic properties of QCD, this model provides us a good background to study various phase structures of strongly coupled field systems. Many efforts have been devoted to investigate properties of this model. Various first and second order hydrodynamic transport coefficients have been calculated in Ref.~\cite{Finazzo:2014cna}. In Refs.~\cite{Janik:2015waa,Janik:2016btb}, quasinormal modes are used to probe the crossover/phase transition and a number of novel features are observed which were not present in the conformal case. In Ref.~\cite{DeWolfe:2010he}, this model is extended to include the effect of finite chemical potential and a more complete phase diagram of QCD is thus studied. Other models are also proposed, see Refs.~\cite{Gursoy:2007cb,Gursoy:2007er,Gursoy:2008za,Kiritsis:2009hu,Gursoy:2009jd} for the improved holographic QCD model (IHQCD), Refs.~\cite{Buchel:2007vy,Buchel:2015saa} for a top-down model and Refs.~\cite{Li:2011hp,Cai:2012xh} for a semi-analytical holographic QCD model.

In this paper, based on Gubser's model, we aim to use one of non-local observables, the entanglement entropy, to probe the crossover/phase transition. Typically, there are three important non-local observables--the two-point function, the Wilson loop and the entanglement entropy--one can consider as probes to track the number of degrees of freedom and reflect non-local information (correlations between parts for example) of the system. However, they are difficult to calculate in the field theory side. AdS/CFT correspondence makes the calculations easier by relating them to some geometric quantities in the bulk. These non-local observables have been extensively used as probes in the study of the holographic thermalization process of strongly coupled field systems (QGP for example). For reviews, see Refs.~\cite{Balasubramanian:2010ce,Balasubramanian:2011ur} and references therein. The time evolution of these non-local observables can reflect thermalization process of different regions of the system and explicitly explain the very short thermalization time. They are also used to describe phase transitions in holographic models of condensed matter systems (holographic superconductor models for example)~\cite{Nishioka:2006gr,Klebanov:2007ws,Pakman:2008ui,Nishioka:2009un,deBoer:2011wk,Hung:2011xb,Ogawa:2011fw,Albash:2011nq,
Myers:2012ed,Albash:2012pd,Cai:2012sk,Cai:2012nm,Cai:2012es,Arias:2012py,Kuang:2014kha,Yao:2014fwa,Dey:2014voa,Ben-Ami:2014gsa,
Peng:2014qfa,Ling:2015dma,Dey:2015ytd,Ling:2016wyr,Peng:2016jor}. In this paper, we will mainly focus on the study of the entanglement entropy, which, according to the AdS/CFT dictionary, is related to the area of some extremal codimension-two surface in the bulk. We will consider three sets of parameters in the scalar potential, which exhibit respectively a crossover, first order and second order phase transitions. Our goal is to investigate the behavior of the entanglement entropy close to the crossover/phase transitions, and to see if this non-local observable can give us some information characterizing the phase structures.

The paper is organized as follows. In the next section, we will give a brief introduction of the holographic QCD model proposed by Guber et al. Then, in Sec. III, thermodynamics of the model is discussed. In Sec. IV, behavior of the entanglement entropy close to the crossover/phase transitions is investigated. We will consider two shapes of the entanglement region: the strip one and the ball one. The last section is devoted to summary and discussions.

\section{Review of the holographic QCD model}

In this section, we give a brief introduction of the holographic model proposed by Gubser et al in Refs.~\cite{Gubser:2008ny,Gubser:2008yx} intending to mimic the equation of state of QCD. The bulk action is a Einstein-dilaton action,
\begin{eqnarray}\label{BulkAction}
S = \int d^5 x \sqrt{-g} \left[R-\frac{1}{2} (\partial \phi)^2 - V(\phi)\right],
\end{eqnarray}
where the scalar potential assumes the form~\cite{Janik:2015waa,Janik:2016btb}
\begin{eqnarray}\label{ScalarPotential}
V(\phi)= -12  \cosh (\gamma \phi) + b_2 \phi^2 + b_4 \phi^4 + b_6 \phi^6.
\end{eqnarray}
This potential is parameterized by four constants, $\gamma, b_2, b_4$ and $b_6$, whose values we can choose. It has the following small $\phi$ expansion
\begin{eqnarray}
V(\phi) \sim  -12 + \frac{1}{2} m^2 \phi^2 + {\cal O} (\phi^4).
\end{eqnarray}
The first term is the negative cosmological constant (note that we have chosen the unit to set the AdS radius to be one), and the second term is the mass term with $m^2 \equiv 2(b_2-6 \gamma^2)$. According to the AdS/CFT dictionary, the scalar field $\phi$ in the bulk is dual to a scalar operator $O_\phi$ in the dual boundary field theory. The conformal dimension of the scalar operator is related to the mass parameter of the scalar field as $\Delta (\Delta -4) =m^2$. The mass square $m^2$ can be negative and is constrained by the Breitenloner-Freedman (BF) bound $m^2 \geq -4$~\cite{Breitenlohner:1982bm,Breitenlohner:1982jf}. Holographically, this gravity model is dual to a deformation of the boundary conformal field theory
\begin{eqnarray}
{\cal L} = {\cal L}_{\rm CFT} + \Lambda^{4-\Delta} O_{\phi},
\end{eqnarray}
where $\Lambda$ is an energy scale. In this paper, we consider $2 \leq \Delta <4$ which corresponds to relevant deformations of the CFT.

By choosing suitable values of the parameters $(\gamma, b_2, b_4, b_6)$, this model can produce an equation of state which agrees well with the lQCD data. Moreover, by choosing other values of the parameters, this model can also realize various types of phase transitions. In this work, as in Refs.~\cite{Janik:2015waa,Janik:2016btb} we consider three sets of parameters, labeled by $V_{\rm QCD}, V_{\rm 1st}$ and $V_{\rm 2nd}$ respectively, which are summarized in Table 1. The parameters for $V_{\rm QCD}$ have been chosen to fit the lQCD data from Ref.~\cite{Borsanyi:2012cr}, and the system is known to possess a crossover behaviour at zero baryon chemical potential as we will show later. Parameters of potentials $V_{\rm 1st}$ and $V_{\rm 2nd}$ were chosen so that the corresponding dual field systems exhibit respectively the $1^{\rm st}$, and the $2^{\rm nd}$ order phase transitions.

\begin{table}[!htbp]
\begin{tabular}{c c c c c c c}
\hline
\hline
potential &  $\gamma$ & $b_2$ & $b_4$ &  $b_6$ &  $\Delta$ \\
\hline
$V_{\rm QCD}$ &  0.606 & 1.4 & -0.1 & 0.0034 & 3.55\\
$V_{\rm 2nd}$ &  $1/\sqrt{2}$ & 1.958 & 0 & 0 & 3.38 \\
$V_{\rm 1st}$ &  $\sqrt{7/12}$ & 2.5 & 0 & 0 & 3.41\\
\hline
\end{tabular}
\caption{Parameters for the three scalar potentials~\cite{Janik:2016btb}.}
\end{table}

As we want to study properties of the dual field system at finite temperature, in the gravity side we need black hole solutions. To seek these solutions, we take the following ansatz as in Refs.~\cite{Gubser:2008ny,Gubser:2008yx},
\begin{eqnarray}\label{ansatz}
ds^2 &=& e^{2 A} (-h dt^2 + d\vec{x}^2) + \frac{e^{2 B}}{h} dr^2.\nonumber\\
\phi &=& r,
\end{eqnarray}
where $A, B$ and $h$ are only functions of $r$ (or, equivalently $\phi$). The above ansatz takes a gauge $\phi=r$ which greatly simplifies the solving of the field equations. Then the field equations of motion are
\begin{eqnarray}
A''-A' B' + \frac{1}{6} &=& 0,\label{FieldEq1}\\
h''+(4 A' - B') h' &=& 0,\label{FieldEq2}\\
6 A' h' + h (24 A'^2 -1) + 2 e^{2B} V &=& 0,\label{FieldEq3}\\
4 A'-B'+\frac{h'}{h} - \frac{e^{2 B}}{h} V' &=& 0,\label{FieldEq4}
\end{eqnarray}
where the prime denotes a derivative with respect to $\phi$. The horizon $\phi=\phi_H$ is determined by the zero point of the blackening function $h$:
\begin{eqnarray}\label{horizon}
h(\phi_H)=0.
\end{eqnarray}
We follow the method proposed in Refs.~\cite{Gubser:2008ny,Gubser:2008yx} to solve the field equations, in which by defining a function $G(\phi) \equiv A'(\phi)$ the solution of field equations can be expressed as:
\begin{eqnarray}
A(\phi) &=& A_H + \int_{\phi_H}^\phi d\tilde{\phi} G(\tilde{\phi}),\\
B(\phi) &=& B_H + \ln\left(\frac{G(\phi)}{G(\phi_H)}\right) + \int_{\phi_H}^{\phi} \frac{d\tilde{\phi}}{6 G(\tilde{\phi})},\\
h(\phi) &=& h_H + h_1 \int_{\phi_H}^{\phi} d\tilde{\phi} e^{-4 A(\tilde{\phi}) + B(\tilde{\phi})},
\end{eqnarray}
where the integration constants $A_H, B_H, h_H$ and $h_1$ are determined by requiring the appropriate boundary conditions at the horizon Eq.~(\ref{horizon}) and the infinite boundary,
\begin{eqnarray}
A_H &=& \frac{\ln \phi_H}{\Delta-4} + \int_0^{\phi_H} d\phi \left[G(\phi) - \frac{1}{(\Delta-4)\phi}\right],\\
B_H &=& \ln \left(-\frac{4 V(\phi_H)}{V(0) V'(\phi_H)}\right) + \int_0^{\phi_H} \frac{d\phi}{6 G(\phi)},\\
h_H &=& 0,\\
h_1 &=& \frac{1}{\int_{\phi_H}^0 d\phi e^{-4 A(\phi) + B(\phi)}}.
\end{eqnarray}
So, once we get the solution of $G(\phi)$, the full solution can be generated. As in Ref.~\cite{Gubser:2008ny}, by manipulating Eqs.~(\ref{FieldEq1})(\ref{FieldEq2})(\ref{FieldEq3})(\ref{FieldEq4}), it is found that $G(\phi)$ satisfies the following "master equation"
\begin{eqnarray}
\frac{G'}{G + V/3 V'}=\frac{d}{d\phi}\ln\left(\frac{G'}{G} + \frac{1}{6 G} - 4 G -\frac{G'}{G + V/3V'}\right).
\end{eqnarray}
From it, the series expansion of $G(\phi)$ near the horizon $\phi=\phi_H$ can be obtained,
\begin{eqnarray}
G(\phi) = -\frac{V(\phi)}{3V'(\phi)} + \frac{1}{6} \left(\frac{V(\phi_H) V''(\phi_H)}{V'(\phi_H)^2}-1\right) (\phi-\phi_H) + {\cal O} (\phi-\phi_H)^2,
\end{eqnarray}
which can be used as the appropriate boundary conditions to solve $G(\phi)$. Note that it is hard to solve the "master equation" analytically, so we rely on numerical method.

From the above expressions, we can see that given one value of the horizon $\phi_H$ we can obtain one unique black hole solution. In this paper, we will vary the value of $\phi_H$ and obtain a family of black hole solutions numerically.

\section{Thermodynamics}

In this section, we study the thermodynamics of the dual field system. We will focus on the temperature-dependence of the entropy density and the speed of sound.

From the ansatz Eq.~(\ref{ansatz}), the Hawking temperature and the entropy density can be obtained,
\begin{eqnarray}
T = \frac{e^{A_H -B_H} |h'(\phi_H)|}{4\pi}, \qquad s = \frac{e^{3 A_H}}{4}.
\end{eqnarray}
From them, we can get the square of the speed of sound
\begin{eqnarray}
c_s^2 = \frac{d \ln T/d\phi_H}{d\ln s/d\phi_H}.
\end{eqnarray}
In the following three subsections, we will discuss the three cases listed in Table 1 respectively.

To make a comparison with the conformal case, here we also show the results for the five-dimensional Schwarzschild-AdS black hole with the metric
\begin{eqnarray}
ds_{\rm SAdS}^2 &=& \frac{1}{z^2} \left[-h(z) dt^2 +\frac{dz^2}{h(z)} + d\vec{x}^2\right],\nonumber\\
h(z) &=& 1 - \left(\frac{z}{z_H}\right)^4.
\end{eqnarray}
The horizon is at $z=z_H$ and the infinite boundary $z=0$. From the above metric and using the AdS/CFT dictionary, it is easy to derive the following relations
\begin{eqnarray}
s/T^3 \propto ({\rm number\ of\ degrees\ of\ freedom}),\qquad c_s^2 = \frac{1}{3},
\end{eqnarray}
which are expected for a CFT.

\subsection{$V_{\rm QCD}$}

In Fig.~1, we show the dependence of the entropy density $s$ and the square of the speed of sound $c_s^2$ on the temperature for $V_{QCD}$. From the right panel, we can see that there is a critical temperature $T_c$ which corresponds to the lowest dip of $c_s^2$. According to Ref.~\cite{Finazzo:2014cna}, this value should be $143.8 MeV$ for QCD, while in our present units it is $T_c=0.181033$. As shown in the left panel, the entropy density and its derivative to the temperature are both continuous at the critical temperature which means a crossover. When the temperature is beyond the critical point, $s/T^3$ approaches constant suggesting that even in the non-conformal case $s/T^3$ may be considered to approximately count the effective number of degrees of freedom. This claim is also true for the other two cases, $V_{\rm 1st}$ and $V_{\rm 2nd}$, as we will see later. Moreover, as the temperature approaching $T_c$, the entropy density $s/T^3$ drops quickly indicating that the number of underlying degrees of freedom is largely suppressed, which can be understood as a signal of confinement. The dependence of $c_s^2$ on the temperature agrees well with the lattice result~\cite{Borsanyi:2012cr,Janik:2015waa}. Moreover, for high temperature, $c_s^2$ approaches its conformal value $1/3$ as expected.

\begin{figure}[!htbp]
\centering
\includegraphics[width=0.45\textwidth]{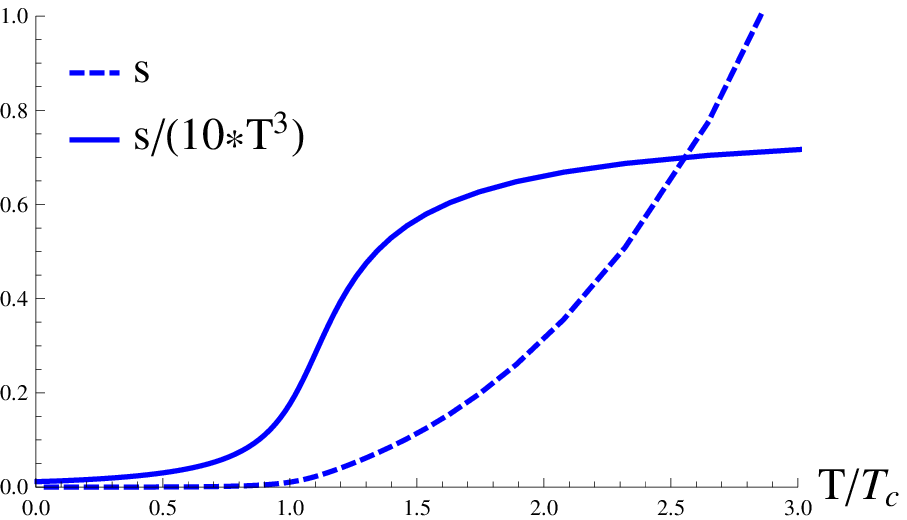}\qquad
\includegraphics[width=0.45\textwidth]{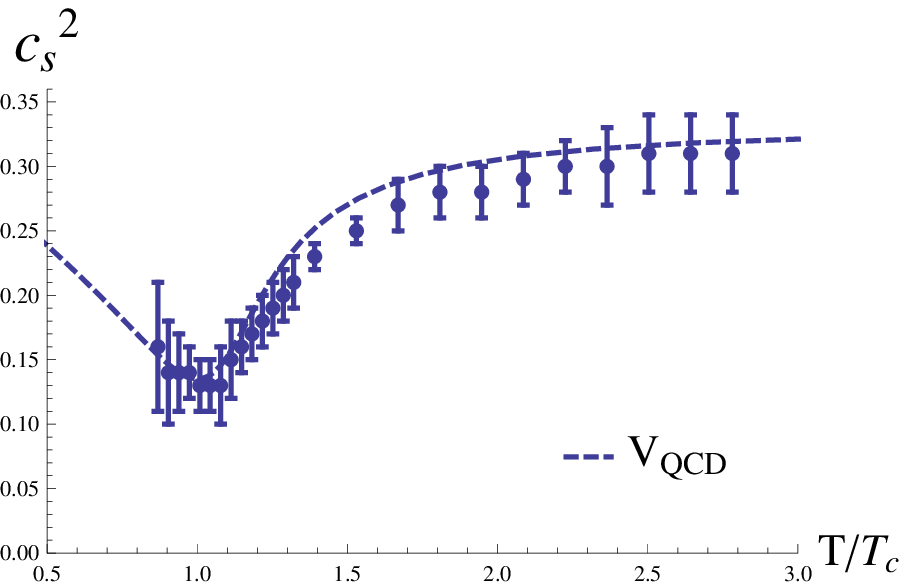}
\caption{(color online) Entropy density $s$ and speed of sound $c_s^2$ for $V_{\rm QCD}$. In the left panel, the two curves are for $s$ (dashed) and $s/(10*T^3)$ (solid) respectively. In the right panel, the lattice data points with error bar are from Ref.~\cite{Borsanyi:2012cr}.}
\end{figure}

\subsection{$V_{\rm 1st}$}

In Fig.~2, the dependence of the entropy density and the square of the speed of sound on the temperature for $V_{1st}$ are shown. From the left panel, we can see that In some range of the temperature, there are three branches of solutions, in which two of them are stable (shown in blue curves) and one is unstable (shown in red curve). And there is a minimum temperature $T_m$ below which no unstable black hole solutions exist. This can be seen more clearly from the right panel, where for the unstable branch of solutions, $c_s^2$ becomes negative. In our present units, $T_m=0.232287$. There is a phase transition between the two stable branches of solutions with critical temperature $T_c \simeq 1.05 T_m$~\cite{Janik:2016btb}, which can be read off by comparing the free energy of the two stable branches of solutions. As the entropy density is discontinuous at the critical temperature, the phase transition is first order. Moreover, from the left panel, we can see that $s/T^3$ drops suddenly as the temperature approaching $T_c$ which also indicates that the number of underlying degrees of freedom is largely suppressed, and can also be understood as a kind of confinement. However, we should note that in the cases $V_{\rm 1st}$ and $V_{\rm 2nd}$, we do not intend to mimic the equation of state of QCD, but rather realize various types of phase structures within the same framework.

\begin{figure}[!htbp]
\centering
\includegraphics[width=0.45\textwidth]{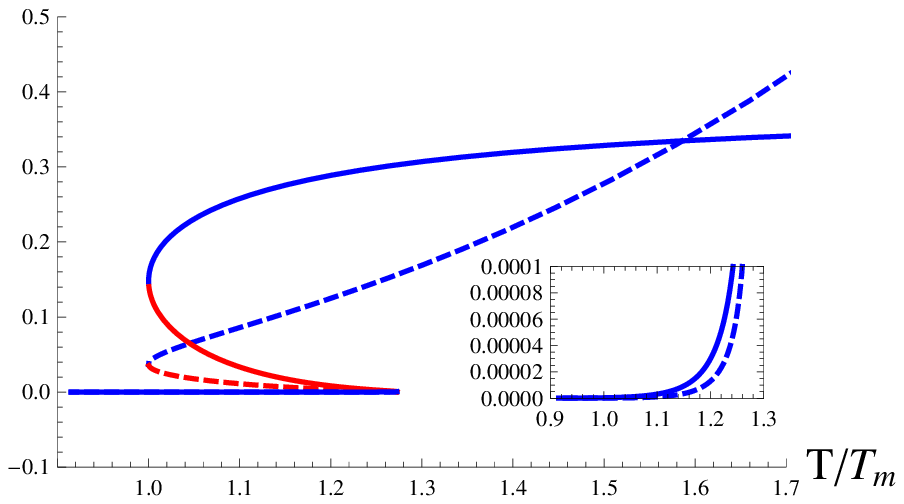}\qquad
\includegraphics[width=0.45\textwidth]{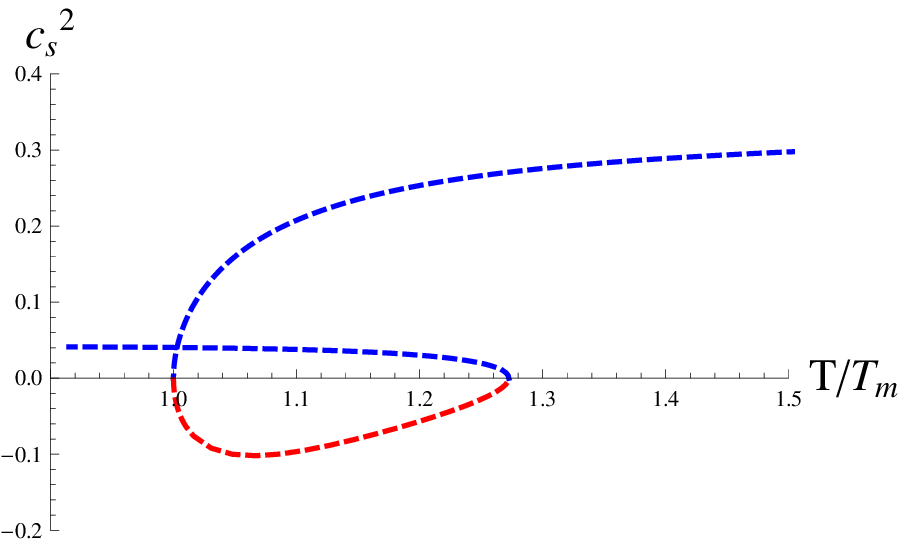}
\caption{(color online) Entropy density $s$ and speed of sound $c_s^2$ for $V_{\rm 1st}$. In the left panel, the two curves are respectively $s$ (dashed) and $s/(20*T^3)$ (solid). In a certain range of temperature, there are three branches of solutions, two of which are stable (blue curves) and one is unstable (red curve).}
\end{figure}

\subsection{$V_{\rm 2nd}$}

In Fig.~3, the dependence of the entropy density and the square of the speed of sound on the temperature for $V_{2nd}$ are shown. From the figure, we can see that there is a critical temperature $T_c$ at which the speed of sound $c_s$ vanishes. The entropy density $s$ is continuous at $T_c$ but not its derivative with respect to the temperature, thus the phase transition is second order. In our present units, $T_c=0.156841$. Once again, we can see that $s/T^3$ drops quickly as the temperature approaching $T_c$.

\begin{figure}[!htbp]
\centering
\includegraphics[width=0.45\textwidth]{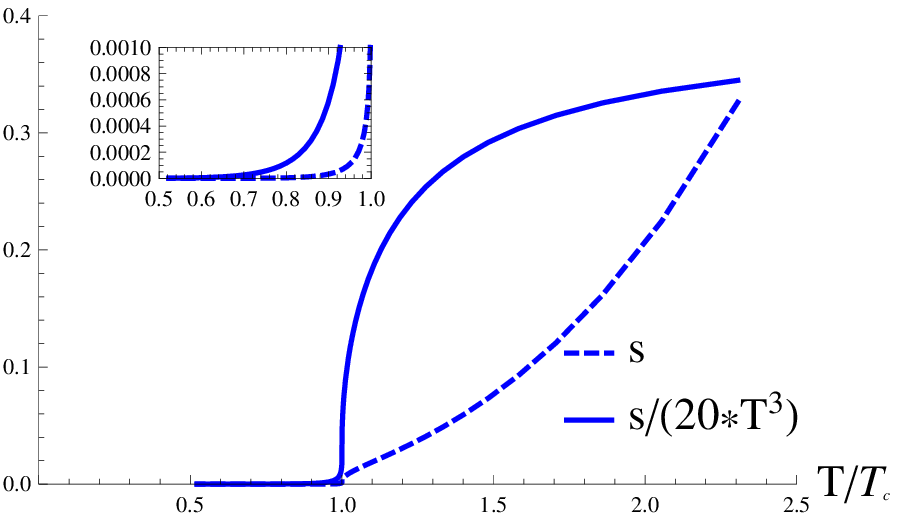}\qquad
\includegraphics[width=0.45\textwidth]{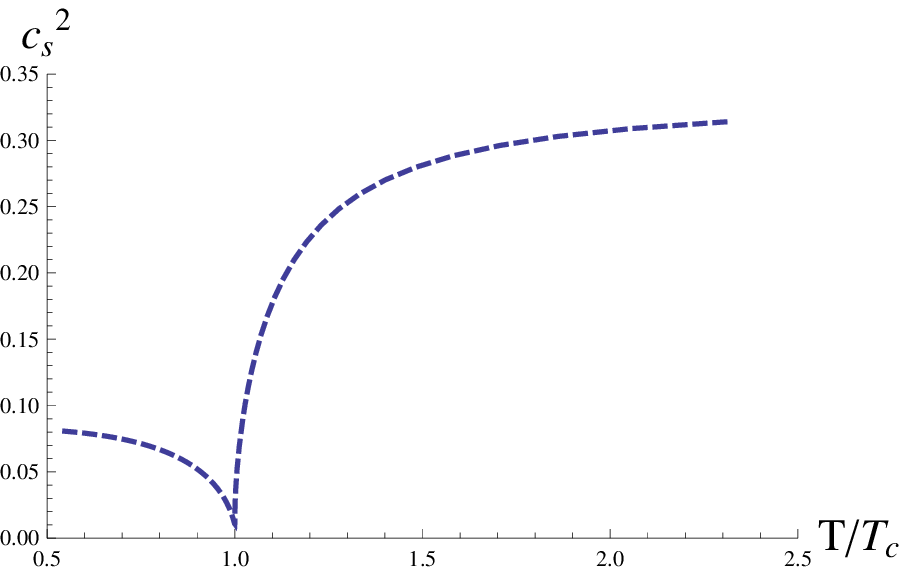}
\caption{(color online) Entropy density $s$ and speed of sound $c_s^2$ for $V_{\rm 2nd}$. In the left panel, the two curves are for $s$ (dashed) and $s/(20*T^3)$ (solid) respectively.}
\end{figure}

\section{Behaviors of holographic entanglement entropy}

Suppose the system of the boundary field system living in Minkowski spacetime is divided into two parts ${\cal A}$ and ${\cal A}^c$, with ${\cal A}^c$ being the complement of ${\cal A}$. Then the entanglement entropy of the subregion ${\cal A}$ is defined as the Von Neuman entropy,
\begin{eqnarray}
S_{\rm EE} \equiv -\mathrm{tr}_{\cal A} \rho_{\cal A} \ln \rho_{\cal A},
\end{eqnarray}
where $\rho_A$ is the reduced density matrix of ${\cal A}$.

Generally, it is difficult to calculate the entanglement entropy directly from the field theory side. However, the calculation can be made easier with the help of the holographic method, which relates this entropy to some geometric quantity of the dual bulk geometry. According to the conjecture proposed in Refs.~\cite{Ryu:2006bv,Hubeny:2007xt}, in Einstein gravity, the holographic entanglement entropy formulae is
\begin{eqnarray}\label{HEE}
S_{\rm EE} = \frac{1}{4 G_N^{(d+1)}} {\rm ext}[{\rm Area}(\gamma_{\cal A})],
\end{eqnarray}
where $\gamma_{\cal A}$ is the extremal codimemsional-two surface in the bulk which shares the same boundary with ${\cal A}$, i.e. $\partial \gamma_{\cal A} = \partial {\cal A}$. $G_N^{(d+1)}$ is the Newton constant which we set to be one in the present work.

In this section, we would like to study the behaviors of entanglement entropy close to the crossover/phase transition, and to see if this non-local observable can characterizes the corresponding phase structures. We will consider two shapes of the entanglement region ${\cal A}$: strip one and ball one.

\subsection{Coordinate redefinition}

Before diving into the calculations of the three non-local observables, we would like to transform the metric into the form as
\begin{eqnarray}\label{NewMetric}
ds^2 = e^{2 \tilde{A}(z)} \left[-\tilde{h}(z) dt^2 + \frac{d z^2}{\tilde{h}(z)} + d\vec{x}^2\right],
\end{eqnarray}
which can be achieved by redefining the radial coordinate as
\begin{eqnarray}
\phi = \phi(z).
\end{eqnarray}
The form of $\phi(z)$ can be obtained by matching the line elements before and after the transformation. The result is
\begin{eqnarray}
z (\phi)= \int_0^\phi d\tilde{\phi} e^{B(\tilde{\phi}) - A(\tilde{\phi})},
\end{eqnarray}
and then $\tilde{A}(z) = A(\phi)$ and $\tilde{h}(z) = h(\phi)$. In this new radial coordinate, the horizon lies at $z=z_H \equiv z(\phi_H)$ and the infinite boundary at $z=0$. We found that it is more convenient to work with the $z$-coordinate to do the calculations. In the following, without causing confusion, we will omit the tilde of the metric functions in Eq.~(\ref{NewMetric}) for the sake of simplicity.

\subsection{Entanglement entropy: Strip shape}

On the boundary with coordinates $(t,\vec{x})$, we consider the entanglement region ${\cal A}$ to be a strip:
\begin{eqnarray}
{\cal A}: \qquad x_1 \in [-\l/2, \l/2],\quad x_2,x_3 \in [-L/2, +L/2],
\end{eqnarray}
which has a width $\l$ in the $x_1$ direction and length $L$ in the other two spatial directions. We consider the case $L \gg \l$ so that ${\cal A}$ preserves translation invariance in the $x_2$ and $x_3$ directions. According to the holographic entanglement entropy formulae Eq.~(\ref{HEE}), the entanglement entropy of the entanglement region ${\cal A}$ can be calculated by computing the area of the extremal surface which starts from the boundary of the region and extended into the bulk. Taking into account of the symmetry, the extremal surface $\gamma_{\cal A}$ can be parameterized by only one function $z=z(x_1)$, and thus the induced metric on the surface is
\begin{eqnarray}
ds^2_{\gamma_{\cal A}} = e^{2 A} \left(1 + \frac{z'^2}{h}\right) dx_1^2 + e^{2 A} (dx_2^2 + dx_3^2).
\end{eqnarray}
The entanglement entropy then is
\begin{eqnarray}
S_{EE} &=& \frac{1}{4} \int dx_1 dx_2 dx_3 \sqrt{\gamma},\nonumber\\
&=&\frac{V_2}{2} \int_0^{\l/2} dx_1 Q^{1/2} e^{3 A},\\
Q &\equiv& 1 + \frac{z'^2}{h},
\end{eqnarray}
where $V_2 \equiv \int dx_2 dx_3$ and $\gamma$ is the determinant of the induced metric on $\gamma_{\cal A}$. The equation for $z(x_1)$ can be obtained by extremizing $S_{EE}$, and the result is
\begin{eqnarray}
2 h z'' - \left(6 h \frac{d A}{d z} + \frac{d h}{d z}\right) z'^2 -6 h^2 \frac{d A}{d z}=0.
\end{eqnarray}
To solve $z(x_1)$, boundary conditions are needed, which, taking into account of the symmetry, are
\begin{eqnarray}
z(0) = z_\ast,\qquad z'(0) = 0,\qquad z(\pm \l/2) =0.
\end{eqnarray}
$z_\ast$ is the deepest position the extremal surface can reach in the $z$-direction.

From the expression of $S_{EE}$, we can see that it does not depend on $x_1$ explicitly which leads to a conservation equation
\begin{eqnarray}
Q^{1/2} = \frac{e^{3 A}}{e^{3 A(z_\ast)}}.
\end{eqnarray}
Using it, $S_{EE}$ can be expressed as
\begin{eqnarray}\label{SEE}
S_{EE} = \frac{V_2}{2} \int_0^{\l/2} dx_1 \frac{e^{6 A(z)}}{e^{3 A(z_\ast)}}.
\end{eqnarray}

To study the behavior of the entanglement entropy close to the crossover/phase transition, we define a renormalized entanglement entropy density $s_{\rm EE}^{\rm re} \equiv \frac{S_{\rm EE} - S_{EE}^0}{\l V_2}$ ($\l V_2$ is the volume of the strip) and study its dependence on the temperature. $S_{\rm EE}^0$ is the entanglement entropy of the strip in some reference spacetime. Generally, the integration in the expression of the holographic entanglement entropy Eq.~(\ref{SEE}) diverges, and by introducing $S_{\rm EE}^0$ the divergence can be renormalized. We choose the reference spacetimes as follows: For the Schwarzschild-AdS case, it is the pure AdS spacetime; For the $V_{\rm QCD}$ and $V_{2nd}$ cases, they are the critical black hole solutions with the critical temperature $T_c$; For the $V_{1st}$ case, it is the critical black hole solution with the temperature $T_m$. Without losing generality and for simplicity, we fix the width of the strip $l=0.04$ and change the temperature while keeping in the regime $T l \ll 1$.

In Fig.~4, the dependence of the renormalized entanglement entropy density on the temperature for the three scalar potentials are shown. To make a comparison, we also show the result for the Schwarzschild-AdS black hole which corresponds to the conformal case.

\begin{figure}[!htbp]
\centering
\subfigure[~~Schwarzchild-AdS]{\includegraphics[width=0.45\textwidth]{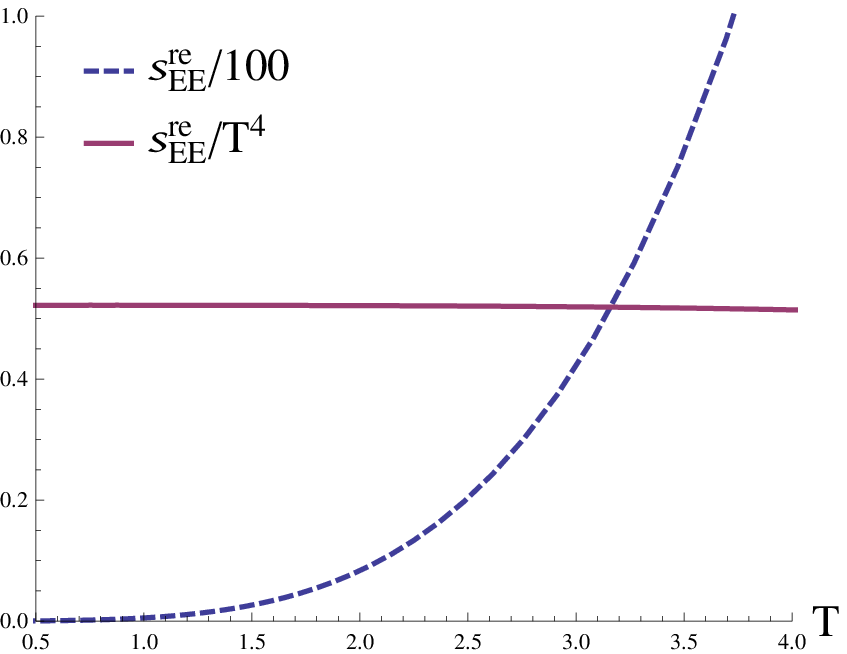}}\qquad
\subfigure[~~$V_{\rm QCD}$]{\includegraphics[width=0.45\textwidth]{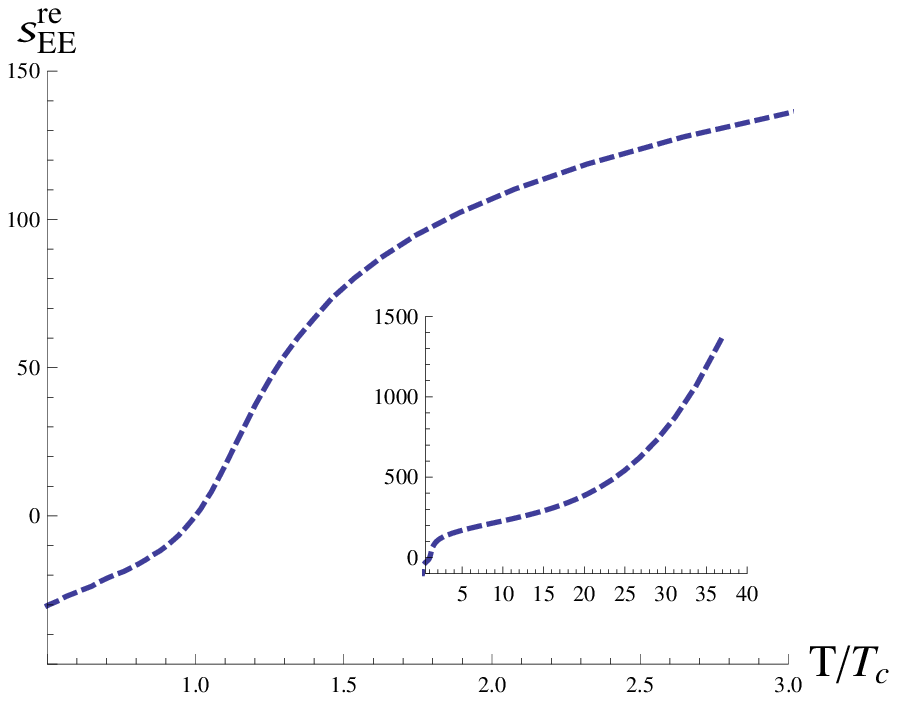}}\qquad
\subfigure[~~$V_{\rm 1st}$]{\includegraphics[width=0.45\textwidth]{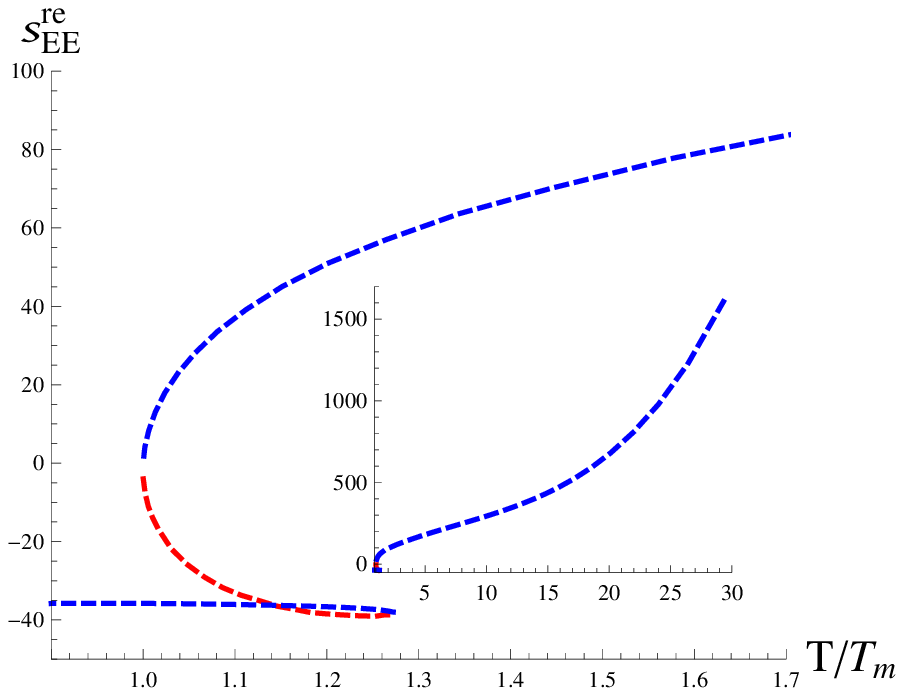}}\qquad
\subfigure[~~$V_{\rm 2nd}$]{\includegraphics[width=0.45\textwidth]{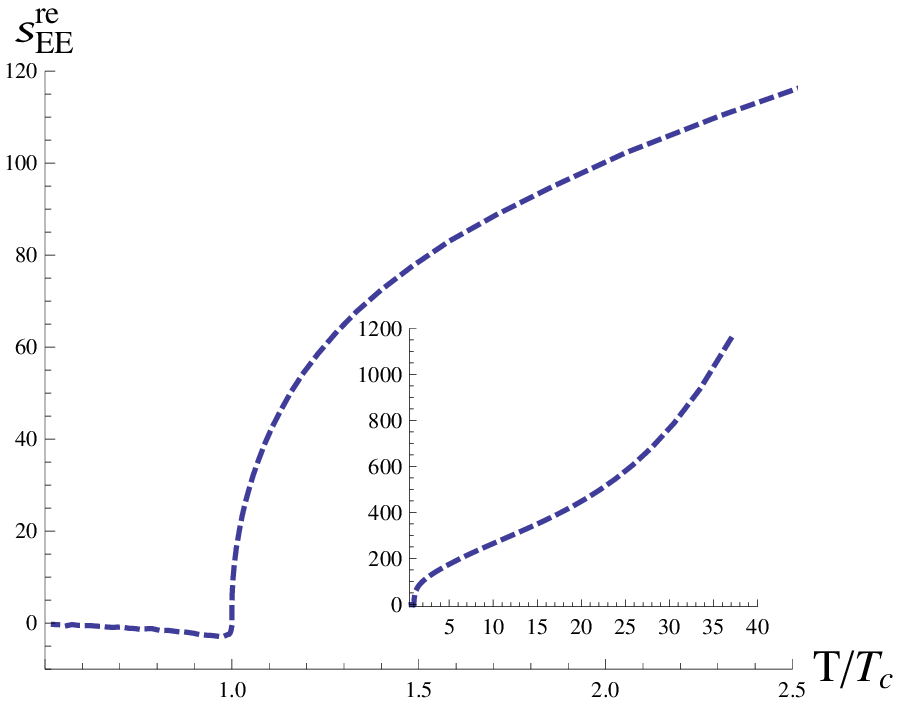}}\qquad
\caption{(color online) Entanglement entropy density for Schwarzschild-AdS black hole solution and the three scalar potentials. The width of the entanglement strip is fixed to be $\l=0.04$. The red curve in the panel (c) corresponds to the unstable branch of solutions. The insets of panels (b),(c) and (d) show the behaviors of renormalized entanglement entropy density at sufficiently high temperature.}
\end{figure}

From the panel (a), we can see that for the Schwarzschild-AdS case, $s_{\rm EE}^{\rm re} /T^4 =const.$. This numerical result agrees with the analytical result in Ref.~\cite{Fischler:2012ca} where it is shown that in the regime $T \l \ll 1$, $s_{\rm EE}^{\rm re} /T^4 \propto N_c^2$ ($N_c$ denotes the color number of the dual conformal field theory and $N_c^2$ is thus the effective number of degrees of freedom). However, from panels (b), (c) and (d), we can see that for the three non-conformal cases, the entanglement entropy shows very different behavior. Different from the observable of the thermal entropy density, here we can not expect $s_{\rm EE}^{\rm re} /T^4$ to count the effective number of degrees of freedom in the non-conformal cases. This difference can be understood as the effect of the additional energy scale $l$ beyond the mass $m$ of the scalar field (or equivalently the conformal dimension $\Delta$ of the scalar operator), which is absent for the observable of thermal entropy density.

We can also see that as $T$ approaching $T_c$, similar to the thermal entropy density, the renormalized entanglement entropy density drops quickly/suddenly. Physically, we can understand this behavior as follows: as the temperature approaching $T_c$, the field system undergoes confinement so that the number of degrees of freedom contributing to the entanglement is largely suppressed. This behavior has also been seen in holographic superconductor models~\cite{Albash:2012pd}, where condensation decreases the entanglement entropy. Moreover, from the insets of panels (b), (c) and (d) and by fitting the data, we can see that at sufficiently high temperature (while keeping in regime $T l \ll 1$), the behaviors of renormalized entanglement entropy density are restored to that of the conformal case, i.e., $s_{\rm EE}^{\rm re} \sim T^4$. This indicates that the effects of conformal symmetry breaking become insignificant at sufficiently high temperature.

Moreover, for the three non-conformal cases, the renormalized entanglement entropy density exhibit behavior characterizing the corresponding phase structures close to $T_c$: For the crossover case $V_{\rm QCD}$, $s_{\rm EE}^{\rm re}$ and its derivative with respect to the temperature are both continuous at $T_c$; For the $1^{\rm st}$ order case $V_{\rm 1st}$, $s_{\rm EE}^{\rm re}$ is discontinuous at $T_c$; For the $2^{\rm nd}$ order case $V_{\rm 2nd}$, $s_{\rm EE}^{\rm re}$ is continuous at $T_c$ but not its derivative with respect to the temperature. These behaviors suggest that, as the thermal entropy, the entanglement entropy may also be used to characterize the type of phase transition.

\subsection{Entanglement entropy: Ball shape}

In this subsection, we consider the entanglement region ${\cal A}$ to take a ball shape with radius $R$, i.e. $\sum_{i=1}^3 x_i^2 \leq R^2$. It is more convenient to work with spherical coordinates $(\rho,\Omega_2)$, under which the bulk metric takes the form
\begin{eqnarray}
ds^2 = -e^{2 A} \left[-h dt^2 + \frac{d z^2}{h} + d\rho^2 + \rho^2 d\Omega_2^2\right].
\end{eqnarray}
Then, in the spherical coordinates $(\rho,\Omega_2)$, the entanglement region is parameterized as $\rho \leq R$. Taking into account of the symmetry, the extremal surface $\gamma_{\cal A}$ can be parameterized by only one function, i.e. $z=z(\rho)$, and with boundary conditions
\begin{eqnarray}\label{BallBoundaryConditions}
z(0) = z_\ast,\quad z'(0) = 0,\quad z(R)=0,
\end{eqnarray}
where $z_{\ast}$ is the tip of the extremal surface denoting the deepest position the extremal surface can reach in the $z$-direction. The induced metric on the extremal surface is
\begin{eqnarray}
ds_{\gamma_{\cal A}}= e^{2 A} \left(1+ \frac{z'^2}{h} \right) d\rho^2 + e^{2 A} \rho^2 d\Omega_2^2.
\end{eqnarray}
The entanglement entropy then is
\begin{eqnarray}
S_{EE} &=& \frac{1}{4} \int d\rho d\Omega_2 \sqrt{\gamma}\\
&=& \pi \int_0^R d\rho \sqrt{1+ \frac{z'^2}{h}} e^{3 A} \rho^2.
\end{eqnarray}
The equation for $z(\rho)$ can be obtained by extremizing $S_{EE}$, and the result is
\begin{eqnarray}\label{BallEq}
2 h \rho z'' + 4 z'^3 - \left(6 h \rho \frac{d A}{d z} + \rho \frac{d h}{d z}\right) z'^2 + 4 h z'-6 h^2 \rho \frac{d A}{d z}=0.
\end{eqnarray}
Note that this equation has a singular point at $\rho =0$, so practically in doing numerical calculations we impose the boundary conditions near the central point $\rho=0$ to avoid numerical problems,
\begin{eqnarray}
z(\epsilon) = z_\ast + {\cal O}(\epsilon^2),\quad z'(\epsilon) = {\cal O}(\epsilon),
\end{eqnarray}
where $\epsilon$ is a small parameter with typical order of $10^{-3}$. The higher order terms can be obtained by solving the equation Eq.~(\ref{BallEq}) near $\rho=0$ order by order.

To study the behavior of the entanglement entropy close to the crossover/phase transition, similarly we define a renormalized entanglement entropy density $s_{\rm EE}^{\rm re} \equiv \frac{S_{\rm EE} - S_{\rm EE}^0}{4\pi R^3/3}$ ($4\pi R^3/3$ is the volume of the entanglement ball) and study its dependence on the temperature. $S_{\rm EE}^0$ is the entanglement entropy of the ball in the reference spacetime with the same choices as in the strip case. Without losing generality and for simplicity, we fix the radius of the entanglement ball to be $R=0.04$ and change the temperature while keeping in the regime $T R \ll 1$. The results are shown in Fig.~5.

\begin{figure}[!htbp]
\centering
\subfigure[~~Schwarzchild-AdS]{\includegraphics[width=0.45\textwidth]{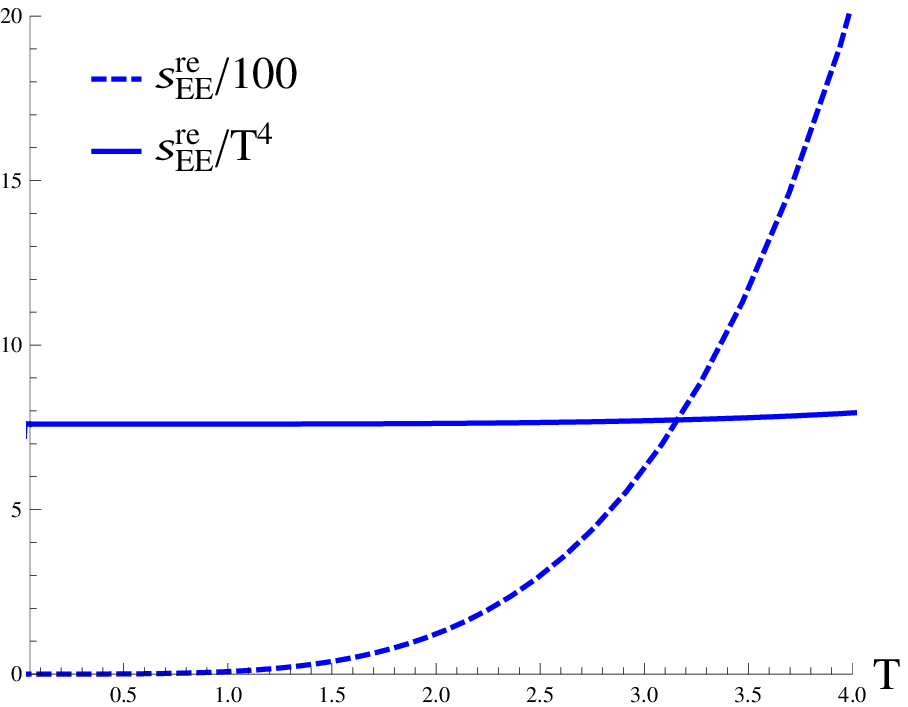}}\qquad
\subfigure[~~$V_{\rm QCD}$]{\includegraphics[width=0.45\textwidth]{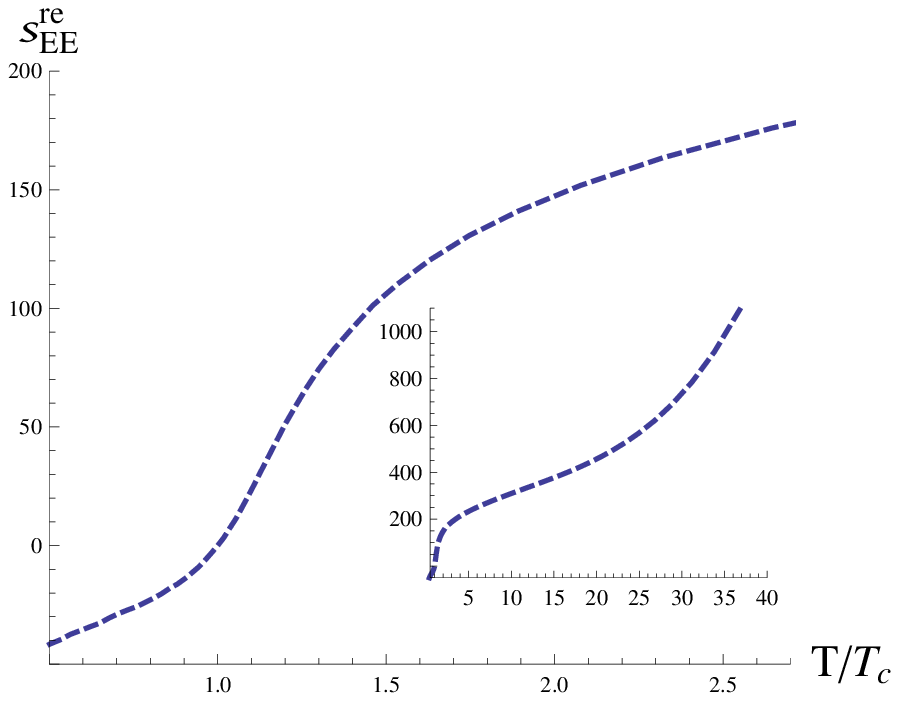}}\qquad
\subfigure[~~$V_{\rm 1st}$]{\includegraphics[width=0.45\textwidth]{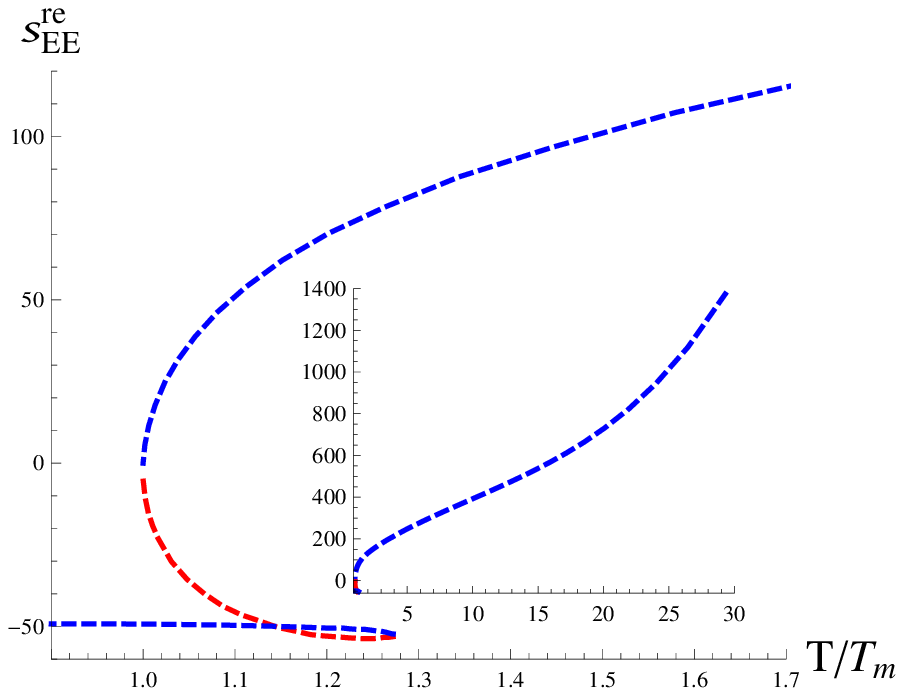}}\qquad
\subfigure[~~$V_{\rm 2nd}$]{\includegraphics[width=0.45\textwidth]{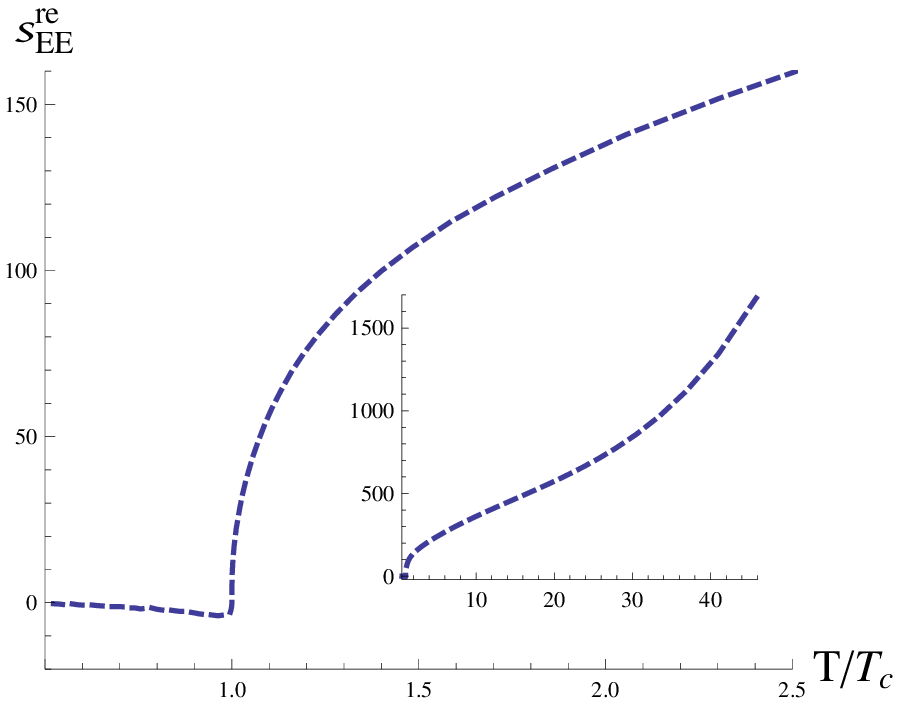}}\qquad
\caption{(color online) Entanglement entropy density for Schwarzschild-AdS black hole solution and the three scalar potentials. The radius of the entanglement ball is fixed to be $R=0.04$. The red curve in the panel (c) corresponds to the unstable branch of solutions.}
\end{figure}

From panel (a), we can see that for the conformal case, in the regime $T R \ll 1$ once again we have the relation $s_{\rm EE}^{\rm re} /T^4 =const.$. For the three non-conformal cases, the renormalized entanglement entropy density shows a similar dependence on the temperature as in the strip case. These results suggest that the behavior of entanglement entropy close to $T_c$ does not depend on the specific shape of the entanglement region. This implies that our previous suggestion, that the entanglement entropy may be applied to characterize the type of the phase transition, does not depend on the specific shape of the entanglement region.

\section{Summary and Discussions}

In this paper, we discuss the behavior of entanglement entropy close to crossover/phase transition in the holographic QCD model proposed by Gubser et al~\cite{Gubser:2008ny,Gubser:2008yx}. This holographic model is proposed intending to mimic the equation of state of QCD by introducing a nontrivial scalar field in the bulk to break the conformal symmetry. The scalar self-interaction potential is parameterized by four constants whose values we can choose to fit the lQCD results. Moreover, by choosing other values of the four parameters, this simple model can also realize various types of phase structures. In this paper, we consider three sets of parameters $V_{\rm QCD}, V_{\rm 1st}$ and $V_{\rm 2nd}$ possessing respectively crossover, $1^{\rm st}$ and $2^{\rm nd}$ phase transitions, which can be seen by studying their thermodynamic properties, such as the entropy and the square of the speed of sound.

Our results show that, similar to the thermal entropy, the entanglement entropy drops quickly/suddenly as the temperature approaching the critical value. This can be understood as a signal of confinement. Moreover, at the critical temperature, it is found that the entanglement entropy shows behavior characterizing the type of the phase transition which is also similar to the thermal entropy. These results suggest that we may apply the entanglement entropy to characterize the phase structures of strongly coupled field systems. Moreover, by studying two cases with different shapes of entanglement region, the strip one and the ball one, we show that our claim does not depend on the specific shape of entanglement region.

We only consider the holographic QCD model proposed by Guber et al~\cite{Gubser:2008ny,Gubser:2008yx}. Whether our claim is still hold for other holographic models, for example the improved holographic QCD model proposed in Refs.~\cite{Gursoy:2007cb,Gursoy:2007er,Gursoy:2008za,Kiritsis:2009hu,Gursoy:2009jd}, the top-down model in Refs.~\cite{Buchel:2007vy,Buchel:2015saa} and for the semi-analytical holographic QCD model in Refs.~\cite{Li:2011hp,Cai:2012xh}, needs further investigations.

\section*{Acknowledgement}

This work was supported in part by National Natural Science Foundation of China (No.11605155).

\end{document}